\documentclass[12pt,preprint]{aastex}

\newcommand{\lsun}{L$_{\odot}$}                   
\newcommand{\msun}{M$_{\odot}$}
\newcommand{\mdot}{$\dot{M}$}
\newcommand{\msunyr}{M$_{\odot} \, {\rm yr}^{-1}$} 
\newcommand{\halpha}{H$\alpha$}                   
\newcommand{\powten}[1]{10$^{#1}$}
\newcommand{\um}{$\mu$m}                          
\newcommand{\asecdot}[2]{\mbox{#1$\stackrel {\prime \prime}{_{\bf \cdot}}$#2}}
\newcommand{\kms}{km\,s$^{-1}$}       

\slugcomment{Submitted to ApJL, last changed 2001-09-21 (AB)}

\shorttitle{The subarcsecond triple star VW\,Cha}
\shortauthors{Brandeker et al.}

\begin{document}

\title{Discovery of a New Companion and Evidence of a Circumprimary Disk:\\
Adaptive Optics Imaging of the Young Multiple System
VW\,Cha
\renewcommand{\thefootnote}{\fnsymbol{footnote}}\footnotemark[1]
\renewcommand{\thefootnote}{\arabic{footnote}}}

\author{Alexis Brandeker\altaffilmark{1,2}, Ren\'e Liseau\altaffilmark{1,2}
and Pawel Artymowicz\altaffilmark{1}}
\affil{Stockholm Observatory}
\email{firstname@astro.su.se}

\and

\author{Ray Jayawardhana\altaffilmark{3}}
\affil{University of California, Berkeley}
\email{rayjay@astron.berkeley.edu}

\renewcommand{\thefootnote}{\fnsymbol{footnote}}
\footnotetext[1]{Based on observations collected at the European Southern
Observatory, Chile}
\renewcommand{\thefootnote}{\arabic{footnote}}

\altaffiltext{1}{Stockholm Observatory, SCFAB, SE-106 91 Stockholm,
Sweden.}
\altaffiltext{2}{Visiting Astronomer, European Southern Observatory ESO, Chile.}
\altaffiltext{3}{Department of Astronomy, University of California,
601 Campbell Hall, Berkeley, CA 94720, U.S.A.}

\begin{abstract}
Since a majority of young low-mass stars are members of multiple
systems, the study of their stellar and disk configurations is crucial
to our understanding of both star and planet formation processes. Here
we present near-infrared adaptive optics observations of the young multiple 
star system VW\,Cha. The previously known \asecdot{0}{7} binary is clearly resolved 
already in our raw $J$ and $K$ band images. We report the discovery of a new, 
faint companion to the secondary, at an apparent separation of only
\asecdot{0}{1} or 16\,AU. Our high-resolution photometric observations
also make it  possible to measure the $J - K$ colors of each of the three
components  individually. We detect an infrared excess in the primary,
consistent with theoretical models of a circumprimary disk. Analytical
and numerical calculations of orbital stability show that VW\,Cha 
may be a stable triple system. Using models for the age and total mass of
the secondary pair, we estimate the orbital period to be 74 years. Thus,
follow-up astrometric observations might yield direct 
dynamical masses within a few years, and constrain evolutionary
models of low-mass stars. 
Our results demonstrate that adaptive optics imaging in conjunction
with deconvolution techniques is a powerful tool for probing close multiple
systems.
\end{abstract}


\section{Introduction}
Most stars are members of binary or multiple systems \citep{bat73, hal83}.
Yet the formation of multiple stellar systems is still poorly
understood. Currently, the most promising mechanism appears to be fragmentation
during the collapse of dense molecular cloud cores \citep{bod00}.
Recent surveys of pre-main-sequence (PMS) stars in certain star-forming 
regions have measured a binary frequency about twice that in the solar 
neighborhood (see \citet{mat00} for a review and caveats). Whether this 
is an evolutionary or environmental effect is open to debate. In any case, 
higher-order multiple systems appear to be relatively rare in both young and 
old stellar populations (see, e.g., \citet{ghe97} and references therein).

High-resolution observations of PMS multiple systems help determine the 
stellar and disk configurations of young objects, leading to better 
constraints on models of multiple star formation and early evolution.
We have undertaken an adaptive optics imaging program of known PMS close 
binary systems to search for additional companions and to detect circumstellar
or circumbinary disks. In this {\it Letter}, we present our observations 
of one such system, VW\,Cha in the Chamaeleon\,I star-forming cloud at a distance
of $\sim$160\,pc \citep{whi97}. The VW\,Cha system has previously been reported to 
consist of several components; \citet{sch91} listed Sz\,23 located $16''$ away 
as a possible wide companion; \citet{bra96} discovered a closer companion at
\asecdot{0}{7} separation; \citet{ghe97} observed these components and reported an additional
companion at \asecdot{2}{7} separation from the primary. \citet[BZ97]{bra97} obtained
spatially resolved spectra of the close binary.

Here, we present the discovery of a new close companion to the secondary of 
VW\,Cha and an infrared excess from the primary in the system, probably due 
to a dusty disk.

\section{Observations}
We observed the VW\,Cha system on 11-13 March 2000
with the European Southern Observatory 3.6-meter telescope at La Silla,
Chile, using the adaptive optics near infrared system (ADONIS) and the
SHARPII+ camera. The SHARPII+ camera is based on a 256$\times$256
NICMOS~III array. We used a plate scale of 0.035\,arcsec/pixel, 
Nyquist-sampling the diffraction limited point spread function (PSF) and
giving an effective field of view of \asecdot{8}{5}$\times$\asecdot{8}{5}.
$J$ band (1.25\,\um)
observations of VW\,Cha were conducted during the first night and $K$ band
(2.18\,\um) during the second. At $J$ we obtained 60 frames,
each with an integration time of 10\,s, chopping the source out of the
field of view for a 10\,s integration on sky between each frame. The
science target itself was used as the wavefront sensor of the
adaptive optics (AO) system. The seeing during the VW\,Cha observations the
first night, as reported by the La Silla seeing monitor, fluctuated in the
$V$ band
(0.55\,\um) between \asecdot{0}{8} and \asecdot{1}{4}. The $K$ band observations of
the second night were obtained in a similar manner with 10 frames of 30\,s
integration time each, with the difference that the sky frames were
obtained after the series of on-source frames and not in between. The
seeing in $V$ this time was stable at about \asecdot{1}{0}. During both
nights most frames showed a diffraction limited core while the Strehl
ratios fluctuated between 0.1 and 0.3.
 
After each series of observations of VW\,Cha, an identical series of
observations with the same AO correction parameters were obtained of a
PSF calibrator star, for post processing purposes. We used the F5V star
SAO\,256804 with $V$ magnitude 8.50.
    
\section{Reductions and Analysis}

The basic data reductions were done in a standard way by subtracting sky
frames from source frames and then dividing by a flat field obtained on
sky during dusk. The resulting frames were then not stacked together but
instead processed with the myopic deconvolution algorithm
IDAC\footnote{Information about IDAC and the package itself may be found
on the web pages of ESO, \url{http://www.eso.org/}}. 

 From the PSF calibrator star observations, we obtained an initial guess of
the PSF during the science data acquisition. Since the PSF calibrator star
data were not obtained simultaneously with the science data, the initial
guess PSF is then modified by IDAC iteratively to better match the
different frames of science data. A series of PSF estimates corresponding
to the science object series is thus produced along with the deconvolved
image. The deconvolution procedure using IDAC took about two weeks of
processing time on an UltraSparc workstation. We refer to \citet{chr99}
for details on IDAC.
 
We performed astrometry of the VW\,Cha system using the deconvolved images
in both $J$ and $K$. The error in separation was estimated by
checking the consistency of measured separations of other multiple star
systems observed several times during the two nights. The error in
position angle was then estimated from the error in separation, by
assuming positional errors to be isotropic.
 
For flux calibration, we used observations of 10 other SAO stars distributed
over the sky and obtained
during the two nights in both $J$ and $K$. Their known $B$ and $V$
magnitudes and spectral classes were used to derive their $J$ and $K$
magnitudes using standard colors \citep{bes88} and an estimate of the
color excess due to extinction using Table~1 of \citet{mat90}, where
we assumed an optical extinction ratio $R_V=3.1$. We then fitted a
relation to the observed count rates and derived magnitudes,
estimating the absolute photometric errors from the statistical error
of the fit. By checking the consistency of the measured flux of stars
observed several times during the two nights we derived the relative
photometric errors.

\section{Results and Discussion}

\subsection{Multiplicity of VW\,Cha}

\begin{figure}
\plotone{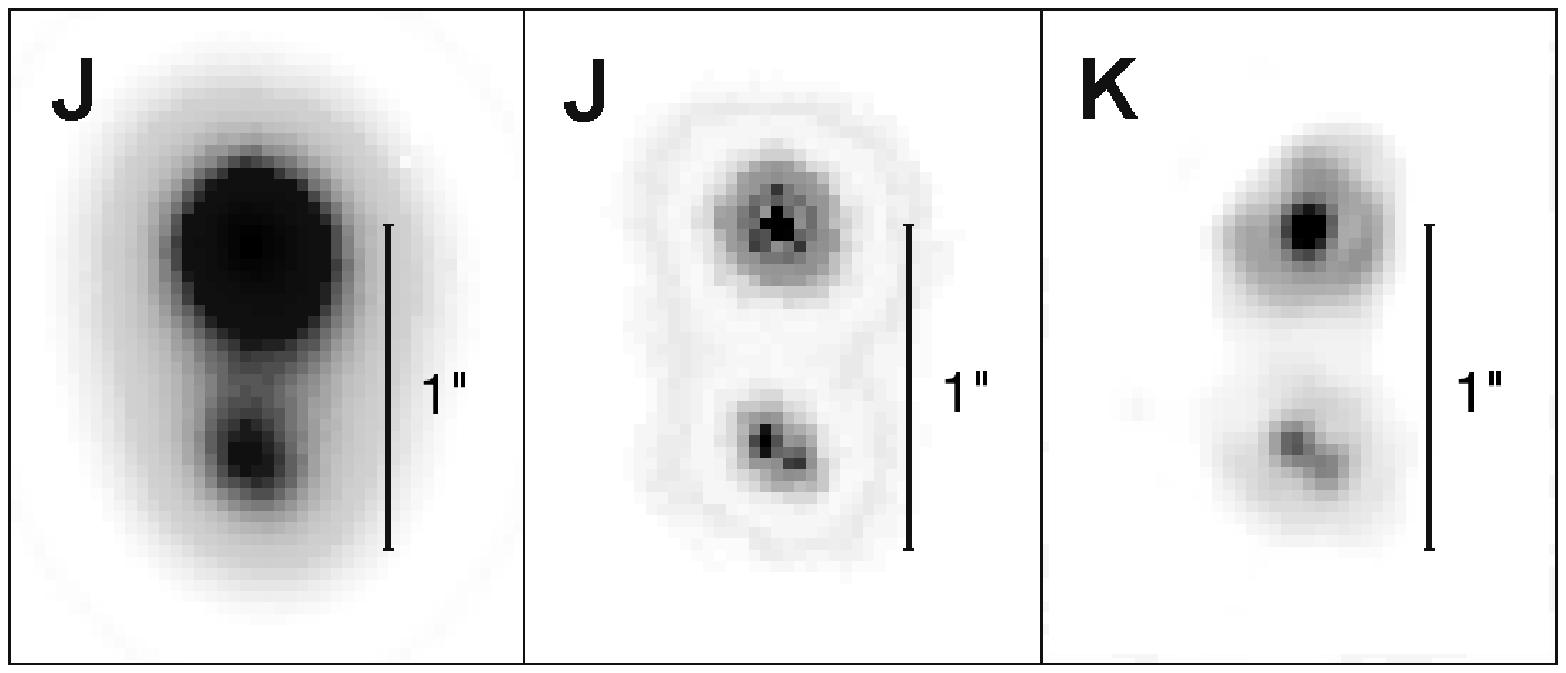}
\caption{An unprocessed frame in $J$ (Left), and deconvolved images in $J$
(Middle) and $K$ (Right) of the VW\,Cha system. The secondary itself is
split into a $0.10''$ binary in the deconvolved images. The vertical bar
in each image corresponds to one arcsecond. North is up, east is to the
left.
\label{fig-1}}
\end{figure}

In our observations, we are able to detect binaries at separations between
\asecdot{0}{08} and $4''$, where the lower limit corresponds to the diffraction
limit of the 3.6-m telescope in $J$, and the upper limit is due to the
\asecdot{8}{5}$\times$\asecdot{8}{5} field of view. The inner \asecdot{0}{7} binary is already
resolved in the raw data in both $J$ and $K$. After deconvolution, the 
secondary is split
up into a tight \asecdot{0}{10} binary in itself; see Fig.~\ref{fig-1}.
Our results are summarized in Table~\ref{tbl-1}.

The reported \asecdot{2}{7} companion was not seen in any of our
images. Based on the background noise levels, our $5\sigma$ detection limits 
are $J=20$ and $K=17$, well below $K \approx 9.4\pm0.6$ derived 
from Table 2 in \citet{ghe97}. The cause for a drop in brightness by a factor 
of $10^3$ is hard
to imagine and we suggest that either the star has drifted out of our field of view,
or that the detection was spurious.

\subsection{Implausibility of chance alignment}

Comparing our observations with measurements from 1992 \citep{bra96}, we find
no significant change (less than \asecdot{0}{05}) of the relative positions
between the A and B+C components. Since VW\,Cha has a proper motion of about
\asecdot{0}{22}\,yr$^{-1}$ \citep{fri98}, the system as a whole has moved at
least $2''$ over the past 10 years. Thus the proper motions of the components
are highly correlated, corresponding to a relative velocity of less than
5\,\kms\ at the distance of VW\,Cha. We therefore conclude that the VW\,Cha
system is not due to chance alignment but a physical triple star.

\subsection{Stellar parameters}

\subsubsection{Corrections for extinction}

 From observations in the visual (BZ97) and in the infrared 
(Table~\ref{tbl-1}) it follows that the primary increasingly dominates the 
emission at increasingly longer wavelengths. Interpolating between the flux
ratios in different wavebands, we aportion the observed total 
$I$-magnitude $=10.60$ \citep{gau92} as 11.03 to A and 11.73 
to B+C. The resulting observed colors,
$(I-J)_{\rm A} = 1.82$ and $(I-J)_{ {\rm B} + {\rm C} } = 1.79$, can then be
compared to the intrinsic colors of K5/K7 and K7 stars (BZ97) for $\log g$
of 3.5 and solar metallicity of the model atmospheres by 
\citet[AHS00]{all00}, viz. ($I-J)_0 = 0.80$ and 0.92 
respectively. The color excess of the relatively `blue star' B+C is 
probably mainly due to foreground dust extinction, hence 
$A_I - A_J = 0.87$ which yields $A_V$\,=3.03\,mag,
where we have used the extinction curve for the Cha\,I cloud 
(G.\,Olofsson, private communication: e.g. $A_V$\,=\,2.90\,$A_J$, 
$A_{ I_{\rm c} } = 1.83\,A_J$, $A_K = 0.37\,A_J$).
The assumption that this value of $A_V$\ also applies to the primary would 
imply excess emission from component A, at the level of about 0.1\,mag at 
$J$ and 0.9\,mag at $K$. An active circumprimary disk would be a natural source
of such excess radiation.

\subsubsection{Stellar luminosities and masses}

For the extinction and excess emission corrected $J$-magnitude of the primary,
$J_{0,\,{\rm A}} = 8.16$, the absolute bolometric magnitude can be obtained
from [AHS00] as $M_{\rm {bol,\,A}} = 3.6$. For B and C, this value is 5.0 
and 5.3, respectively. Consequently, the {\it bolometric} luminosities 
are $L_{\rm A} = 2.9$\,\lsun, $L_{\rm B} = 0.79$\,\lsun\ and 
$L_{\rm C} = 0.59$\,\lsun. The `blue' components B and C contribute to 
the total flux mainly at shorter wavelengths. The neglect of correcting for
extinction is expected, therefore, to underestimate the total luminosity of
the system. This is indeed the case, as the {\it calorimetric} luminosity 
of $L_{\rm cal} = 3$\,\lsun\ by \citet{pru92} constitutes merely 70\% of 
$L_{\rm bol} = 4.3$\,\lsun. A natural explanation of this deficit would be that
only a fraction of the stellar photons are reprocessed to the infrared, while
the rest escapes non-isotropically due to the presence of a circumstellar disk.

The empirically determined parameters of the individual stellar components 
of VW\,Cha are consistent with those of the theoretical evolution models 
(for single stars)\footnote{The validity of these models for binaries has
recently been assessed by \citet{pal01}.} by \citet{pal99} for masses of A,
B and C of 1.0\,\msun, 0.40\,\msun\ and 0.35\,\msun, respectively, at the common
age of $0.4\times 10^{6}$\,yr (Table~\ref{tbl-1}). This assumes a systematic
shift of $\Delta log T_{eff} = -0.04$, not unreasonable.

BZ97 found the \halpha-line to be in emission toward both A and B+C. If dominated
by disk accretion processes, some infrared excess could be expected also from the
fainter pair and should thus be corrected for. However, the \halpha\ flux is larger
from A (BZ97) and the mass accretion rate, \mdot, is presumably also larger for 
the primary. Estimates of \mdot, which are in reasonable agreement with 
the $K$-excess,  are in the range of a few times \powten{-8}\,\msunyr\ 
\citep{gah95} to \powten{-7}\,\msunyr\ \citep{har98}, but certainly much 
less than the large rates (\powten{-6} to \powten{-4}\,\msunyr) obtained by 
\citet{joh00} in the far-UV, a spectral region which is highly susceptible 
to extinction corrections.

VW\,Cha was detected by ISO at 6.7\,\um\ and 14.3\,\um\ \citep{per00}, but not
by IRAS at 60\,\um\ and 100\,\um\ \citep{pru92},
suggesting that only small amounts of dust exist in the system which are 
at temperatures significantly below 50\,K. For the stellar properties and 
mass accretion rate determined above, an outer circumprimary disk radius  
$<70$\,AU could then be inferred from standard theory of steady state, 
thin-disk accretion \citep{fra85}, using a temperature profile of the 
form $T_{\rm disk}(r)\propto r^{-0.4}$. This minimum temperature would 
also be consistent with the scenario of a tidally truncated disk around 
A of size of at most twice the $\overline {\rm BC}$ binary separation, 
advocated in the next section (the separation of \asecdot{0}{2} corresponds
to 30 AU, where $T_{\rm disk}\sim 70$\,K). Further, the observed excess in 
$J$ would be consistent with the presence of hot dust and thus of 
only a small central hole. However, as illustrated by the interpretation of
the observed weakness of the 10\,\um\ silicate feature in VW\,Cha 
\citep{nat00} some caution seems to be called for, when applying 
simplified disk theories developed for single stars to close multiples.

It is interesting to note that the secondary in the T\,Tau binary
system recently has been resolved into a binary in itself \citep{kor00}.
The projected relative distances for A:(B+C) and B:C are about 100\,AU and
10\,AU respectively, making the configuration very similar to VW\,Cha.
In addition, evidence for a circumprimary disk around T\,Tau with a radius 
$\sim$40\,AU has been presented by \citet{ake98}. The major difference is the 
much higher luminosity of the T\,Tau system, $\sim$30\,\lsun\ \citep{coh79}.
An apparent 
difference is also the extremely red appearence of T\,Tau\,S. This has 
previously been interpreted as the (unresolved) secondary being in a 
considerably earlier stage of evolution than the primary. However, later 
results suggest that the IR-excess may be due to foreground extinction (see 
discussion in \citet{ake98}).

\subsection{Triple star dynamics: Stability estimates}

We consider the possibility
that VW\,Cha is a physical triple, bound and stable for
$\sim$1\,Myr (upper limit on age). 
Components B and C form a tight pair B:C at the projected separation of
16\,AU, while A is found at a 7 times larger projected separation of
106\,AU. We first show
analytically that a nearly planar 3-body system similar to VW\,Cha, with low
eccentricities, is forever stable against the disintegration of the close
binary by the third massive body. 

We utilize energetic constraints on the motion of a 3-body system, 
similar to the familiar zero-velocity curves in the restricted 3-body problem
where the Jacobi integral entails 
the so-called Hill stability, involving no exchange of components
and no 'ionization'. A small binary B:C will never be pulled apart by  
component A, if (\citet{mar82}, \citet{mar88})
\begin{equation}
(-E L^2) > (IU^2)_{L1}\, ,
\end{equation} 
where $E$ is the total mechanical energy of the triple system, $L$ its 
total angular momentum, $I$ the semi-moment of inertia (one-half times
the sum of the squares of radius vectors times
masses), and $U$ the total potential energy of the triple.
Subscript $L1$ denotes the value at the collinear ``Lagrange point''
between the two most massive bodies. Stability does not depend on 
either the total mass or the distance units.
E.g., tilting a given system 
with respect to the plane of the sky merely rescales the
inferred periods but not the shapes of orbits. 
If  projection effects are not extreme, the orbital 
periods are close to those in a planar system 
with semi-major axes of $a_{1}=110$\,AU and $a_{2}=16$\,AU for the A:(B+C) 
and B:C pairs, i.e.\ on the order of 872\,yr and 74\,yr. 

For the masses given in Table~\ref{tbl-1}, we found that 
right hand side of the stability criterion equals
$(IU^2)_{L1}=0.0152 GM_{tot}^2$ (where $M_{tot}=1.75$\,\msun\ 
is the total mass). Our results show that if the initial orbit of the
outer binary A:(B+C) is circular ($e_1=0$), then the B:C pair is Hill stable
for a wide range of $e_2$ (its initial eccentricity) 
$e_2<0.37$ and its semi-major axis $a_{2}<26$\,AU.
If $e_1=e_2=0$, stability can be assures for  $0<a_{2}<27$\,AU.
Alternatively, assuming cautiously that in the current configuration
both orbits are at apocenters (maximum extension)
we find that eccentricities $e_1<0.19$ and $e_2<0.20$ are required for 
stability. In summary, the 3-body, $(-E L^2)$ criterion guarantees
that a nearly coplanar hierarchical system (with geometry consistent with 
VW\,Cha), having small eccentricities, will never restructure by ejecting
any star to infinity or form a close pair of either A:B or A:C. 

The analytical criterion cannot guarantee that the B and C
components will not physically collide. It might also indicate instability
where none actually occurs on interesting, short, time scale (which we fix at
1 Myr). To confirm its predictions, we therefore integrated 
the motion of VW\,Cha-like configurations with a
7th-8th order Runge-Kutta method, accurate to 10 significant digits 
in the total energy. We assumed $a_1=110$ AU
and $a_2=16$ AU. Only the ratio $a_1/a_2 \approx 7:1$, 
is decisive for stability. Mutual approaches of stars to less than $10^7$\,km 
($\sim  10$ PMS stellar
radii) were treated as physical mergers following a tidal capture.
We found that the analytical and numerical criteria approximately coincide. 
For instance, we confirmed numerically our analytical predictions discussed
above. 

\begin{figure}
\plotone{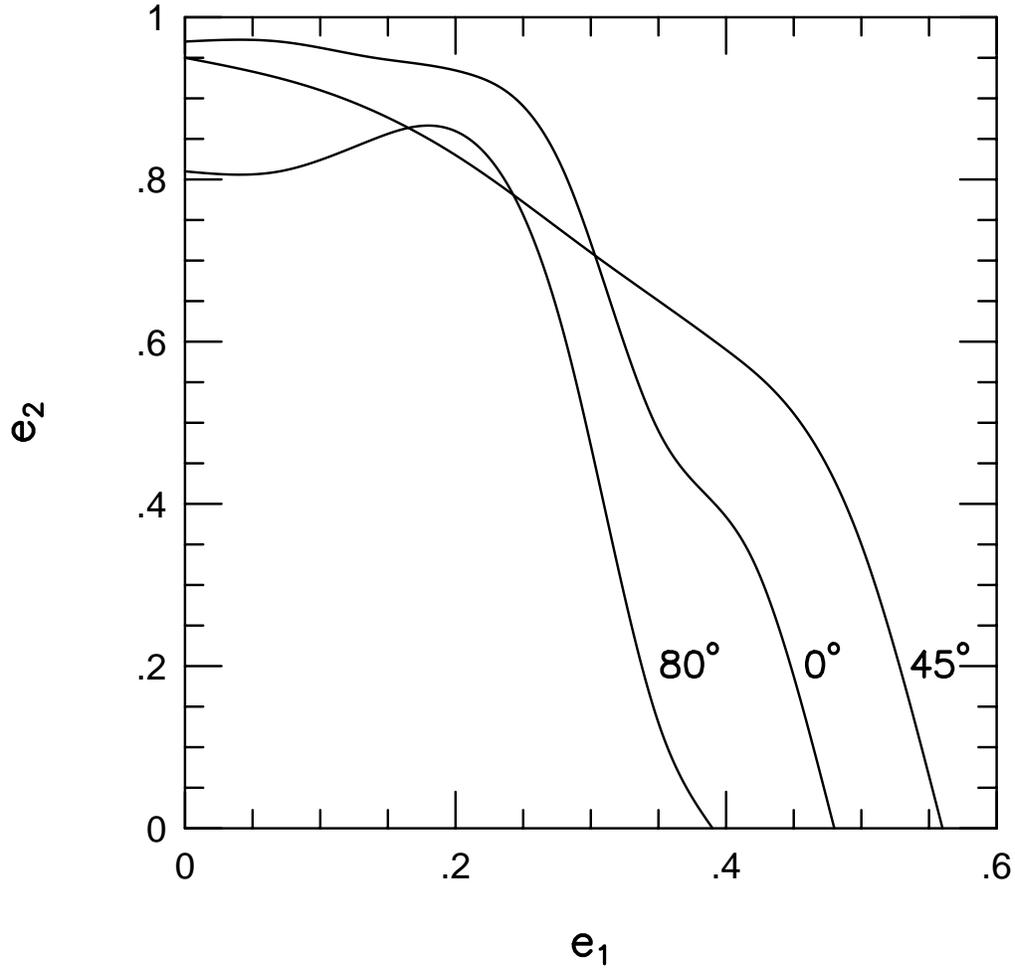}
\caption{Stability diagram of a model triple system representing VW\,Cha.
Eccentricities of a large outer and a small inner binary subsystem are
denoted $e_1$, $e_2$, and their relative inclination is marked next to the curves.
The system is stable at least 1\,Myr for parameter sets below the plotted curves.
\label{fig-2}}
\end{figure}

Fig.~\ref{fig-2} presents numerical results for the planar and inclined 
system configurations. Curves are labeled with 
the starting initial inclination of the inner and outer orbits. 
Combinations of the eccentricities located below the plotted curves 
represent stable systems. All destruction modes including
physical collisions of stars were observed above the curves. 
Strictly speaking, the lines shown in the figure result from 
averaging of the results over such unessential initial parameters
as the longitudes of pericenters, ascending nodes and true anomalies

 From Fig.~\ref{fig-2} we conclude that a wide variety of eccentricities
result
in stable behavior. While $e_1 < 0.45$ is generally required, larger  $e_2$
values (up to 0.8) are consistent with stability, independent of the mutual
orbital inclination (with exception of unstable, almost perpendicular orbits
not presented here). We conclude that the hierachical projected structure of
VW\,Cha likely reflects a stable physical configuration for the estimated age
of the system.

\section{Conclusions}

The young double star VW\,Cha (\asecdot{0}{7} separation) has
been successfully observed with adaptive optics techniques in the 
infrared $J$ and $K$ bands. These AO observations led to the followig main
findings:
\begin{itemize}
\item[$\bullet$] The
secondary is a very close binary itself (\asecdot{0}{1} separation).
\item[$\bullet$] This physical triple system is shown to be stable over
 a period exceeding its estimated age of 0.4\,Myr.
\item[$\bullet$] Follow-up astrometric observations 
may, within a few years, constrain the dynamical mass of the B:C subsystem, 
thus testing evolutionary models of PMS stars.
\item[$\bullet$] The primary (A) has a significant IR-excess, 
indicating the presence of a circumprimary disk. 
This disk is probably small, with a radius not exceding $\sim$30\,AU.
\end{itemize}

Our VW\,Cha result shows that individual components of known binaries may be
close binaries themselves, which needs to be  considered  while deriving ages
and other parameters of such systems.

\acknowledgments
We had many helpful discussions on VW\,Cha with G\"{o}ran Olofsson and
G\"{o}sta Gahm. We are also grateful to Keith Hege for extensive support
with the use of IDAC and to Wolfgang Brandner for supplying photometric
data.

\clearpage

\clearpage

\begin{deluxetable}{crrrrrrrrr}
\tabletypesize{\scriptsize}
\tablecaption{IR Photometry and Astrometry of the VW\,Cha system 2000\,03\,11-13 \label{tbl-1}}
\tablewidth{0pt}
\tablehead{
\colhead{Stellar} & 
\colhead{[$J$] 1.25\,\um\tablenotemark{a}} &
\colhead{[$K$] 2.18\,\um\tablenotemark{b}} &
\colhead{$J-K$} &
\colhead{Separation\tablenotemark{c}} & 
\colhead{Position Angle\tablenotemark{c}} & 
\colhead{$M_{\rm bol}$} &
\colhead{$L$} &
\colhead{$M$} &
\colhead{Age}
\\
\colhead{component} & 
\colhead{(mag)} &
\colhead{(mag)} &
\colhead{(mag)} &
\colhead{($''$)} &
\colhead{($\arcdeg$)} & 
\colhead{(mag)} &
\colhead{(\lsun)} &
\colhead{(\msun)} &
\colhead{(Myr)}	  
}

\startdata
A         &9.21  &7.03 &$2.18\pm0.16$ &\nodata         &\nodata       &3.56 &2.93 &1.00 &0.4    \\
B         &10.55 &8.92 &$1.63\pm0.16$ &$0.661\pm0.006$ &$176.6\pm0.5$ &4.97 &0.79 &0.40 &0.4    \\
C         &10.87 &9.32 &$1.55\pm0.16$ &$0.100\pm0.006$ &$233.3\pm3.3$ &5.29 &0.59 &0.35 &0.4    \\
B+C     &9.94  &8.35 &$1.59\pm0.20$ &$0.683\pm0.008$\tablenotemark{d} &$179.4\pm0.7$\tablenotemark{d}
 &4.37 &1.38 &0.75 &\nodata \\
A+B+C &8.76  &6.75 &$2.01\pm0.24$ &\nodata         &\nodata       &3.14 &4.30 &1.75 &\nodata \\
\enddata

\tablenotetext{a}{Absolute error $\sigma_J = 0.06$\,mag, 
mean relative error $\sigma_J = 0.026$\,mag.}

\tablenotetext{b}{Absolute error $\sigma_K = 0.08$\,mag, 
mean relative error $\sigma_K = 0.12$\,mag.}

\tablenotetext{c}{Separation and position angle for components B and B$+$C
are relative component A, and component C is relative component B. Position
angle is measured from north to east.}

\tablenotetext{d}{For B+C the astrometry is made with respect to the photocenter.
The larger error is due to the shift of the photocenter with wavelength.}

\end{deluxetable}

\end{document}